\documentclass[aps,12pt]{revtex4}
\usepackage{epsfig}
\usepackage{amsmath}
\usepackage{subfigure}
\usepackage{ulem}
\usepackage{amsfonts}
\usepackage{epsfig,bm}
\usepackage{graphicx}
\usepackage{comment}
\newcommand{\be}{\begin{equation}}
\newcommand{\ee}{\end{equation}}
\newcommand{\ba}{\begin{eqnarray}}
\newcommand{\ea}{\end{eqnarray}}
\newcommand{\nn}{\nonumber}

\begin{document}

\title{Scalar Fields in BTZ Black Hole Spacetime and Entanglement Entropy}
\vspace{0.50cm}

\author{Dharm Veer Singh}
\email{veerdsingh@gmail.com}
\author{Sanjay Siwach} \email{sksiwach@hotmail.com}

\affiliation{Department of Physics,
Banaras Hindu University,
Varanasi- 221 005, India.\\}

\vspace{2.5cm}
\noindent
\begin{abstract}
\begin{center}
ABSTRACT
\end{center}
We study  the quantum scalar fields in background of BTZ black hole spacetime. We calculate the entanglement entropy using the discretized model, which resembles a system of coupled harmonic oscillators. The leading term of the entropy formula is standard Bakenstein-Hawking entropy and sub-leading corresponds to quantum corrections to black hole entropy. We calculate the co-efficent of sub-leading logarithmic corrections numerically.   
\end{abstract}

\maketitle
\newpage
\section{Introduction}
Black holes are classical solutions of Einstein's equations in general theory of relativity. They are characterized by few parameters viz. mass, charge and angular momentum. The gravitational pull of black hole is so strong that even light can not escape from it. This creates a problem as everything goes inside and nothing escapes, so no information can be retained about the objects going inside. A quantum mechanical treatment of scalar fields in the black hole background gives rise to Hawking radiation, which is thermal in nature and carries no information. A full quantum theory of gravity for example string theory is needed to resolve this information loss paradox.

The horizon of black hole is a null surface and it is known that area of black hole horizon never decreases in a physical process. This has an analogy with entropy of a thermodynamic system and it was suggested by Bekenstein that black holes obey the laws similar to thermodynamics and their entropy being proportional to the area of black hole horizon. This has been generalized to include the entropy of matter fields in black hole vicinity and the sum of black hole entropy and matter entropy never decreases in any physical process.

There are several attempts  to understand the microscopic origin of black hole entropy and its relation with horizon area. A first principle calculation of entanglement entropy may be the key to understand this puzzle. We carry this investigation for BTZ black hole \cite{Banados:1992,Carlip:1995}. BTZ black hole is a solution of (2+1) dimensional gravity with negative cosmological constant. The Penrose diagram of the BTZ black hole is similar to the Kerr-Newmann AdS black hole in (3+1)-dimension. The entropy is also proportional to the area as  the case of Schwarzschild and Kerr black holes. This solution can be embedded in string theory and a microscopic counting of degrees of freedom responsible for black hole entropy is also known \cite{Strominger:1996,Kim}. The near horizon geometry of BTZ black hole is anti-de-Sitter space time and machinery of AdS/CFT correspondence can also be applied \cite{Birmingham:2001}. Thus BTZ black hole can be used as a useful laboratory to study the quantum aspects of black holes.

In this paper we shall investigate the entropy of BTZ black hole using ab initio calculations of entanglement entropy of BTZ black hole. We study the entanglement entropy of quantum fields in BTZ black hole spacetime \cite{Bombelli,Banados:1992,Srednicki:1993,Carlip:1995,Frolov:1995,Setare:2004,Takayanagi:2006,Casini:2009,Carlip:95,Solodukhin:2011}.  We review a simple quantum mechanical model of a system of simple harmonic oscillators to illustrate the procedure following the work of Bombelli et el \cite{Bombelli}. This is applied to scalar fields in BTZ black hole spacetime and corresponding entanglement entropy is evaluated. We extend our results to estimate the logarithmic corrections  to the black hole  entropy. These terms for  entropy  are sub-leading in $r_+/a$ and are of the form; $c_2\ln\,(r_+/a)+c_3$. We have calculated the numerical values  of these coefficients and compare our result with that of Mann and Solodukhin \cite{MS}.

\section{Model of Entanglement entropy}
Let us consider a system of coupled harmonic oscillators $q^{A}$ $(A=1,.....,N)$ described by the Lagrangian \cite{Bombelli,Mukohyama:1998A,Mukohyama:1998C,Cardy:2008},
\be
{\cal{L}}=\frac{a}{2}\delta_{AB} \dot{q}^{A}\dot{q}^{B}-\frac{1}{2}V_{AB}q^{A}q^{B}.
\ee
Here $\delta_{AB}$ is Kronecker delta symbol and V is real symmetric, positive definite matrix which does not depend upon $q^{A} $ and \lq\lq a\rq\rq  is fundamental length scale characterizing the system.

The corresponding Hamiltonian of the system becomes,
\be
H_{tot}=\frac{1}{2a}\delta^{AB}p_{A}p_{B}+\frac{1}{2}V_{AB}q^{A}q^{B},
\ee
where $p_{A}=a\delta_{AB}\dot{q}^{B}$ is the canonical momentum conjugate to $q^{A}$.

The total Hamiltonian can be rewritten as \cite{Bombelli},
\be
H_{tot}=\frac{1}{2a}\delta^{AB}(p_{A}+iW_{AC}q^{C})(p_{B}-iW_{BD}q^{D})+\frac{1}{2a}TrW
\ee
where $W$ is symmetric, positive definite matrix satisfying the condition $V_{AB}=W_{AC}W^C_B$. $(p_{A}+iW_{AC}q^{C})$ and $(p_{B}-iW_{BD}q^{D})$ are the creation and annihilation operators similar to harmonic oscillator problem and they obey the commutation relation
\be
[a_A,a_B^{\dagger}]=2W_{AB}.
\ee
If $\psi_{0}$ is the ground state for the system, then  $\psi_{0}$ shall obey the condition,
\be
(p_{A}-iW_{AC}q^{C})|\psi_{0}>=0 .
\ee
The solution of equation (5) is given by, \cite{Bombelli}
\ba
\psi_0(\{q^C\})&=&<\{q^C\}|\psi_0>\nn\\
&=&\Big[\det\frac{W}{\pi}\Big] exp (-\frac{1}{2}W_{AB}(\,q^A\,q^B+q^Aq^B)).
\ea
The density matrix is defined is $\rho=|\psi><\psi|$. Substituting the value of $\psi$, the density matrix can be written as;
\ba
\rho(\{q^A\},\{q^B\})&=&<\{q^A\}|\psi_0><|\psi_0|\{q^B\}>\nn\\
&&\Big[\det\frac{W}{\pi}\Big] exp (-\frac{1}{2}W_{AB}\,q^A\,q^B).
\ea
It is useful to split ${q^A}$ into two subsystems, ${{q^a}}$ $(a=1,2,.......n_B)$ and ${q^\alpha}$ $(\alpha=n_B+1,n_B+2,.......N)$.  The matrix W naturally splits into four blocks as,
\[ (W)_{AB} = \left(\begin{array}{ccc}
A_{ab} & B_{a\beta} \\
B^T_{\alpha b} & D_{\alpha \beta}\end{array} \right).\]

The reduced density matrix of the system '1' is obtained by tracing the degrees of freedom of the system '2' and is given by;
\ba
\rho_{red}(\{q^a\},\{q^{b\prime}\})&=&\int \Pi dq^c\langle\{q^a,q^{\alpha}\}|\rho_{0}|\{q^{\prime b},q^{\beta}\}\nn\\
&=&(det\frac{M}{\pi})^{1/2}exp[-\frac{1}{2}M_{ab}(q^aq^b +q'^{a}q'^{b})] [-\frac{1}{4}(N)_{ab}(q-q')^a (q-q')^b]
\ea
where $M_{ab}=(A-BD^{-1}B^{T})_{ab}$ and $N_{ab}=(B^TA^{-1}B)_{ab}$.
 
Now, the system can be diagonalized using the unitary matrix $U$ and the transformations;
\be
q^a\rightarrow \tilde{q}^a=(UM^{1/2})^a_bq^b.
\ee 
The density matrix reduces to the form \cite{Bombelli}, 
\be
\rho_{red}(\{q^a\},\{q^b\})=\Pi_{n}\{\pi^{-1/2}\exp\Big[-\frac{1}{2}(q_n q^n+q^{\prime}_n q^{\prime n}-\frac{1}{4}\lambda_i (q-q^{\prime})_n (q-q^{\prime})^n)\Big]\},
\ee
where $\lambda_i$  are the eigenvalues of the matrix,
\be
\Lambda^a_b=(M^{-1})^{ac}N_{cb}.
\ee 
 
The entropy $S_{ent}=-Tr(\rho_{red}~\ln\rho_{red})$ of the system is given by;
\be
S=-\sum_i \ln(\frac{1}{2}\lambda_i^{1/2})+(1+\lambda_i)^{1/2}\ln[(1+\lambda_i^{-1}+\lambda_i^{-1/2}],
\ee
which can be rewritten using the variables $\mu_i:=\lambda_{i}^{-1}(\sqrt{1+\lambda_{i}}-1)^2$: 
 \be
S_{i}=-\frac{\mu_i}{1-\mu_i}ln\,\mu_i-ln\,(1-\mu_{i})
\ee
The total entropy of the system is $ S=\sum_{i=1}^{N-n_{B}}S_{i}$, where the sum  corresponds the different eigenvalues.

\section{Entanglement Entropy of BTZ Black Hole}
BTZ black hole is solution of (2+1) dimensional gravity with negative cosmological constant \cite{Banados:1992,Kim}. The action of the (2+1) dimensional gravity with cosmological constant can be written as,
\be
S=\frac{1}{2\pi}\int d^3x\sqrt{-g}\,[R+2\Lambda].
\ee
where $\Lambda={-\frac{1}{l^2}}$ is the cosmological constant.  

The metric of BTZ black hole can be written as;
\be
{ds}^2=-(-M+{r}^2/{l}^2){dt}^2+\frac{1}{(-M+{r}^2/{l}^2+{J}^2/{4r}^2)}{dr}^2+{r}^2{d\phi}^2-J{dt}{d\phi}~,
\ee
with $-\infty < t < \infty$ and $0\leq\phi\leq2\pi$.

The  inner  and outer horizon of the black hole are located at
\be
r_{\pm}=l\;\Big[\frac{M}{2}\Big( 1{\pm}\sqrt{1-(J/Ml)^2}\Big )\Big]^{1/2}.\nn
\ee
where the parameters $M$ and $J$ denote the mass and angular momentum of the black hole and they obey, $M>0$ and $|J|< Ml$. In the extremal case $J=Ml$, and both roots coincide.

The proper length from the horizon, $\rho$, is given by, 
\be
r^2=r_+^2\cosh^2{\rho}+r_-^2\sinh^2{\rho}.
\ee
The metric of the black hole can be written in terms of proper length,  
\be
ds^2=-u^2dt^2+d\rho^2 +l^2\,(u^2+M)d\phi^2-Jdt\,d\phi,
\ee 
where we are using, $r^2(\rho)=l^2(u^2+M)$. 

The  action  of massless scalar field in the background of BTZ black hole space time is given by;
\be
S=-\frac{1}{2}\int dt\,\sqrt{-g}\,\,(g^{\mu\nu}\,(\partial_{\mu}\Phi\,\partial_{\nu}\Phi)).
\ee
Using the separation of variable as appropriate for the cylindrical symmetry of the system;
\be
\Phi(t,\rho,\phi)=\sum_{m}\,\Phi_{m}(t,\rho) \,e^{im\phi},
\ee
the action of the scalar field in the background of BTZ black hole can be written as 
\ba
S&=&-\frac{1}{2}\int dt\Big[-\frac{\sqrt{(u^2+M)}}{\sqrt{[u^2+\frac{J^2}{4(u^2+M)}]}}\,\dot{\Phi_m}^2+u\sqrt{[(u^2+M)+\frac{J^2}{4u^2}]}(\partial_\rho\Phi_m)^2\nn\\&+&\frac{u^2m^2}{u\sqrt{[(u^2+M)+\frac{J^2}{4u^2}]}}{\Phi_m}^2-\frac{(iJm)}{u\sqrt{[(u^2+M)+\frac{J^2}{4u^2}]}}\Phi_m\dot{\Phi_m}\Big],
\ea
The conjugate momentum corresponding to $\Phi_m$, is given by
\ba
\pi_m=\frac{\sqrt{(M+u^2 )}}{\sqrt{[u^2+\frac{J^2}{4(u^2+M)}]}}\,\dot{\Phi_m}+\frac{iJm}{2u\sqrt{[(u^2+M)+\frac{J^2}{4u^2}]}}\,{\Phi_m}\nn.
\ea
To diagonalise the Hamiltonian, we define the new momentum
 \be
\tilde{\pi_m}=\pi_m-\frac{iJm}{u\sqrt{[(u^2+M)+\frac{J^2}{4u^2}]}}\Phi_m.
\ee
The Hamiltonian of the system can be written as,
\be
H=\frac{1}{2}\int d\rho\,\tilde{\pi}_{m}^2(\rho)+\frac{1}{2}\int\,d\rho\,d\rho'\psi_m(\rho)V_m(\rho,\rho')\psi_m(\rho'),
\ee
where,
\ba
\psi_{m}(\rho)V_{m}(\rho,\rho')\psi_{m}(\rho')&=&u\sqrt{[(u^2+M)+\frac{J^2}{4u^2}]}\,\Big(\partial_{\rho}(\sqrt{\frac{\sqrt{[u^2+\frac{J^2}{4(u^2+M)}]}}{\sqrt{(u^2+M)}}})\,\psi_m\Big)^2\nn\\&+&m^2{\frac{{u^2+\frac{J^2}{4(u^2+M)}}}{{(M+u^2 )}}}\psi_m^2,
\ea
and we have used the redefined variable $\psi$
\be
\Psi_{m}(t,\rho)=\sqrt{\frac{\sqrt{[u^2+\frac{J^2}{4(u^2+M)}]}}{\sqrt{(M+u^2 )}}}\,\Phi_m(t,\rho).
\ee

The system can be discretized as,
\ba
&&\rho\rightarrow (A-1/2)\,a, \qquad \qquad \delta(\rho-\rho ')\rightarrow \delta_{AB}/a.
\ea
where A,B=1,2 ....N and \lq\lq{}a\rq\rq{} is UV cut-off length. We regain the continuum limit by taking $a \rightarrow 0$ and $N\rightarrow \infty$ while keeping the size of the system fixed. The parameter $a$ does not seem to have any physical significance here, but we shall see later that finiteness of entropy requires the existence of a cut-off length.   

The corresponding Hamiltonian of the discretized system can be obtained by the replacement:
\ba
&&\psi_m(\rho)\rightarrow q^A, \qquad \qquad \tilde{\pi}_m(\rho)\rightarrow p_{A}/a, \qquad \qquad
V(\rho,\rho')\rightarrow V_{AB}/a^2.
\ea
and is given by the expression,  
\be
H_{0}=\sum_m H_0^m =\sum_{A,B=1}^N \Big[\frac{1}{2a}\delta^{AB}p_{mA}\,p_{mB}+\frac{1}{2}V_{AB}^m\,q_m^A\,q_m^B\Big].
\ee
The $N \times N$ matrix representation of the $V_{AB}^m$ is given by 
\begin{eqnarray}
 \left( V^{(m)}_{AB}\right) & = & \left( 
               \begin{array}{cccccc}
                   \Sigma^{(m)}_1 & \Delta_1 & & & & \\
         \Delta_1 & \Sigma^{(m)}_2 & \Delta_2 & & & \\
       & \ddots & \ddots & \ddots & & \\
       & & \Delta_{A-1} & \Sigma^{(m)}_A & \Delta_A & \\
       & & & \ddots & \ddots & \ddots 
               \end{array}
             \right), \nonumber
\label{eqn:Vlm}
\end{eqnarray}
where
\ba
\Sigma_A^{(m)}&=&\frac{\sqrt{[u^2_{A}+\frac{J^2}{4(u^2_{A}+M)}]}}{\sqrt{(u^2_{A}+M)}}\Big(u_{A+1/2}\sqrt{[(u^2_{A+1/2}+M)+\frac{J^2}{4u^2_{A+1/2}}}\nn\\&-&u_{A-1/2}\sqrt{[(u^2_{A-1/2}+M)+\frac{J^2}{4u^2_{A-1/2}}}\Big)+m^2{\frac{{u^2_A+\frac{J^2}{4(u^2_A+M)}}}{{(M+u^2_A )}}}.
\ea
\ba
\Delta_A&=&-u_{A+1/2}\sqrt{[(u^2_{A+1/2}+M)+\frac{J^2}{4u^2_{A+1/2}}}\sqrt{\frac{\sqrt{[u^2_{A+1}+\frac{J^2}{4(u^2_{A+1}+M)}]}}{\sqrt{(u^2_{A+1}+M)}}}\sqrt{\frac{\sqrt{[u^2_{A}+\frac{J^2}{4(u^2_{A}+M)}]}}{\sqrt{(u^2_{A}+M)}}}\nn\\
\ea
The off-diagonal term of the matrix represents the interactions. We compare this with the corresponding Hamiltonian in previous section and evaluate entanglement entropy of the system. The entanglement entropy of the black hole is given by \cite{Huerta:2012},
\be
\label{entp}
S_{ent}=\lim_{N\rightarrow\infty}\,S(n_B,N)=S_0+2\sum_{m=1}^{\infty}\,\,S^m_{ent}
\ee
Here $S_0$ is the entanglement entropy of the system at m=0, $S(n_B,N)$ is the entanglement entropy of the total system $N$ with partition $n_B$, and $S^m_{ent}$ is the entanglement entropy  of the subsystems for a given value of  azimuthal quantum number 'm'. The equation (\ref{entp}) is an infinite series, but it converges faster at the large value of m and we truncate the series whenever the subsequent terms agree upto three decimal place.

\section{Numerical Estimation of Entropy} 
In this section, we calculate the entanglement entropy of BTZ black hole by considering the formalism outlined in previous sections. The entropy of the BTZ black hole is proportional to the area of the horizon, ($2 \pi r_+$);
\be
S_{ent}= c_s\, \frac{r_+}{a},
\ee
where $c_s$ is the numerical constant of order one and we shall estimate this coefficient numerically and $a$ is the UV cut-off used to discretize the system.

To calculate the entanglement entropy, we divide the total system by a hyper-surface $\Sigma$ defined as $R=n_B a$, where the integer $n_B$ specifies the size of the partition. The entropy scales with the size of the system and we expect it to be finite. The only way it can be possible is that entropy is a function of the dimensionless ratio $R/a$. This requires the existence of the a fundamental length scale $a$ in the system. This cut-off can be identified with Planck length in fundamental theories of gravity. 

We plot entropy as a function of $R/a$ for different values of $n_B$ in figure (1) and the plot confirms that entropy varies linearly as a function of $R/a$ within the limits of numerical accuracy. A detailed numerical analysis using very small size lattice is required in order to prove the independence of entropy on the cut-off length, but the results here are indicate that the dependence is through the ratio $R/a$. If the hyper-surface is taken at the horizon of the black hole, then the entropy becomes proportional to $r_+/a$ and we recover the area law of black hole physics. 

In addition to this, we may also expect that the numerical constant $c_s$ will also depend on any other parameter, if any, through this cut-off. In the continuum limit this constant should become independent of these factors. It has been argued that the constant is a slowly varying function of its arguments \cite{Bombelli,Srednicki:1993}.

We calculate the entropy of the BTZ black hole at large N (large N means,by increasing the value of N, the change in entropy is not significant). First we calculate the $S_m$ for different value of $n_B$ at fixed N, and after that the sum over 'm' is performed. As we go towards the higher value of 'm', the error  decreases sharply. The numerical values of various terms in entropy  are listed in table I. The sum of the series is truncated at a point from where the  enough accuracy in leading term (slope of line) upto three decimal points is achieved. 

We plot entropy as a function of $r_+/a$, where $r_+$ is the horizon of the black hole and this gives the value $c_s=.294$ (slope of line in figure (2)). The dependence of entropy on angular momentum $J$ is also shown in the figure (3). The entropy has no explicit dependence on angular momentum except through the factor $r_+/a$, which defines the size of the horizon. 
\begin{center}
\begin{table}[]
\begin{center}
\begin{tabular}{|l|l|r|l|r|l|r|}
\hline
\multicolumn{1}{|c|}{ $n$ } & \multicolumn{1}{c|}{ $S_0(n)$ } & \multicolumn{1}{c|}{ $S_1(n)$ }& \multicolumn{1}{c|}{$S_2(n)$} &\multicolumn{1}{c|}{ } &  & \multicolumn{1}{c|}{$S_{total}(n)$} \\
\hline
\,\, 10\,\, &  \,\, 0.49450\,\, &0.37468 &  \,\,  0.27642 & \,\,\,-\,-\,-\,-\,-\,-\,\,\,& \,\,\,-\,-\,-\,-\,-\,-\,\,\,&2.7764\,\,
 \\
\,\, 20\,\, \,\, \,\, &\,\,  0.59772\,\, & \,\,0.48157\,\,& \,\,  0.38431\,\, &\,\,\,-\,-\,-\,-\,-\,-\,\,\,& \,\,\,-\,-\,-\,-\,-\,-\,\,\,&\,\, 5.1843\,\,
\\
\,\, 30\,\, &\, \,\,0.65119\,\, &0.54109 &   \,\, 0.44881\,\, &\,\,\,-\,-\,-\,-\,-\,-\,\,\, & \,\,\,-\,-\,-\,-\,-\,-\,\,\,& 8.5227\,\,\\
\,\, 40\,\, &\,\, 0.68795\,\, & 0.57730 & \,\,  0.49351 \,\, &\,\,\,-\,-\,-\,-\,-\,-\,\,\,&\,\,\,-\,-\,-\,-\,-\,-\,\,\,&11.5968\,\,
\\
\,\, 50\,\, &\,\, 0.71089\,\, &  0.60557&  \,\,  0.52521\,\, &\,\,\,-\,-\,-\,-\,-\,-\,\,\,&\,\,\,-\,-\,-\,-\,-\,-\,\,\,& 14.6386\,\,\\
\,\, 60\,\, &\,\,  0.72547\,\, & 0.63439 &  \,\, 0.54964  \,\, &\,\,\,-\,-\,-\,-\,-\,-\,\,\,&\,\,\,-\,-\,-\,-\,-\,-\,\,\, &17.6262\,\,
 \\
\,\, 70\,\, &\,\, 0.73391\,\, & 0.65861 &  \,\, 0.56189\,\, &\,\,\,-\,-\,-\,-\,-\,-\,\,\,&\,\,\,-\,-\,-\,-\,-\,-\,\,\,&20.5947 \,\,
\\
\,\, 80\,\, &\,\,  0.73697\,\, & 0.67468 &  \,\, 0.59023 \,\, &\,\,\,-\,-\,-\,-\,-\,-\,\,\,&\,\,\,-\,-\,-\,-\,-\,-\,\,\,& 23.5625\,\,
\\
\,\, 90\,\, &\,\,  0.74502\,\, & 0.68418 &  \,\, 0.61938 \,\, &\,\,\,-\,-\,-\,-\,-\,-\,\,\,&\,\,\,-\,-\,-\,-\,-\,-\,\,\,& 26.5346\,\,
\\
\,\, 100\,\, &\,\,  0.75306\,\, & 0.68712 &  \,\, 0.62971 \,\, &\,\,\,-\,-\,-\,-\,-\,-\,\,\,&\,\,\,-\,-\,-\,-\,-\,-\,\,\,&\,\, 29.4987\,\,\\
 \hline
\end{tabular}
\end{center}
\caption{For the fixed value of N=100, we calculate the entropy for the  value of $n_B$ and "m" is taken in S.}
\label{tab:high}
\end{table}
\end{center} 

We can extend our results to estimate the logarithmic corrections to black hole entropy. We write the entropy in the form of, 
\be
S_{log}=c_1\,(r_+/a)+c_2\,\log(r_+/a)+c_3,
\ee
and fitted the data points and found the numerical value of these coefficients , $c_1=.303,\qquad c_2=-.104,\qquad c_3=-.186$.  

The leading term of this equation Bekenstein Hawking entropy and the sub-leading term describe the first quantum correction due to quantum entanglement. The co-efficent of sub-leading term can be compared with that of Mann and Solodukhin \cite{MS}, where it is found to be ($-\frac{1}{6}$).
\begin{figure}
\centering
\includegraphics[width=.75\textwidth]{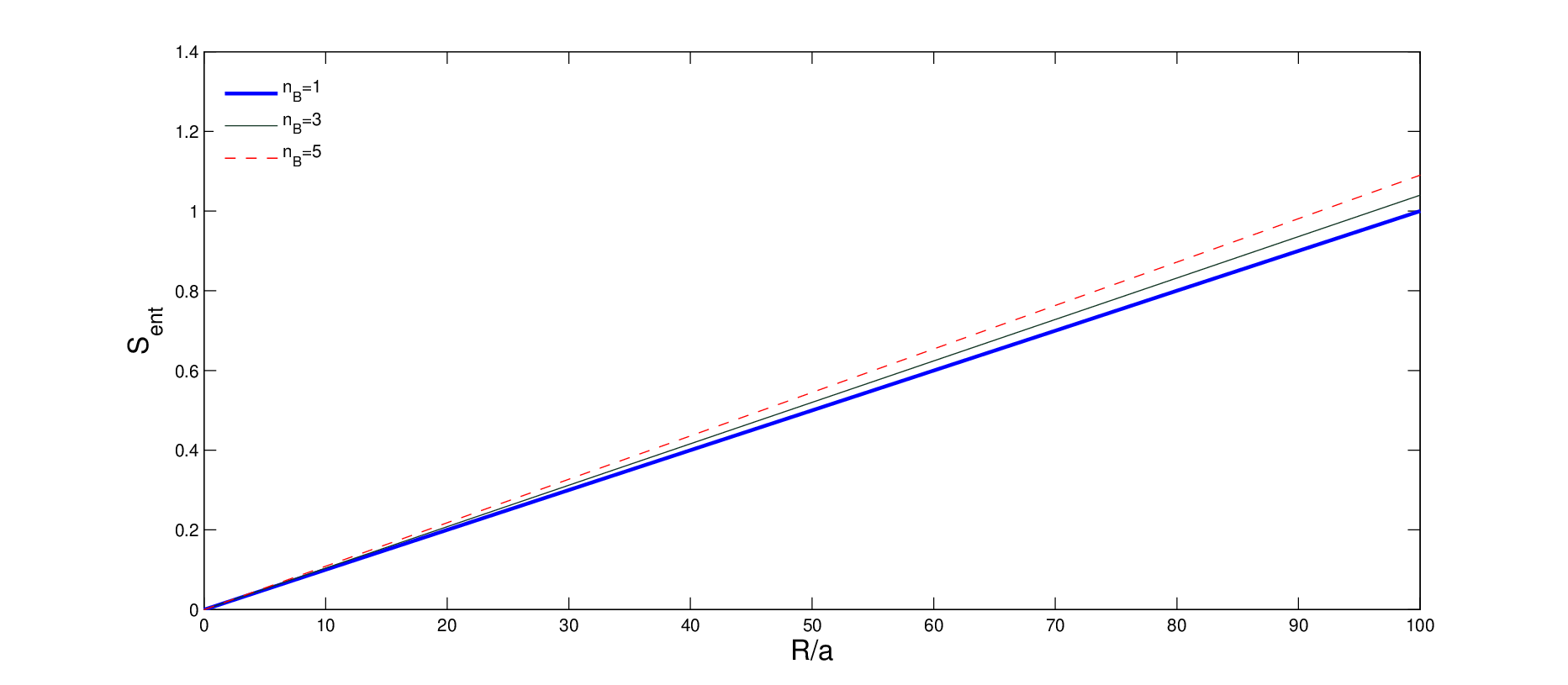}
\caption{The numerical calculation for $S_{ent}$ of the scalar field in the BTZ spacetime. $S_{ent}$ for $n_B=1,3,5$ as a function $R/a$ for N=100}
\label{the-label-for-cross-referencing}
\end{figure}
\begin{figure}
\centering
\includegraphics[width=.75\textwidth]{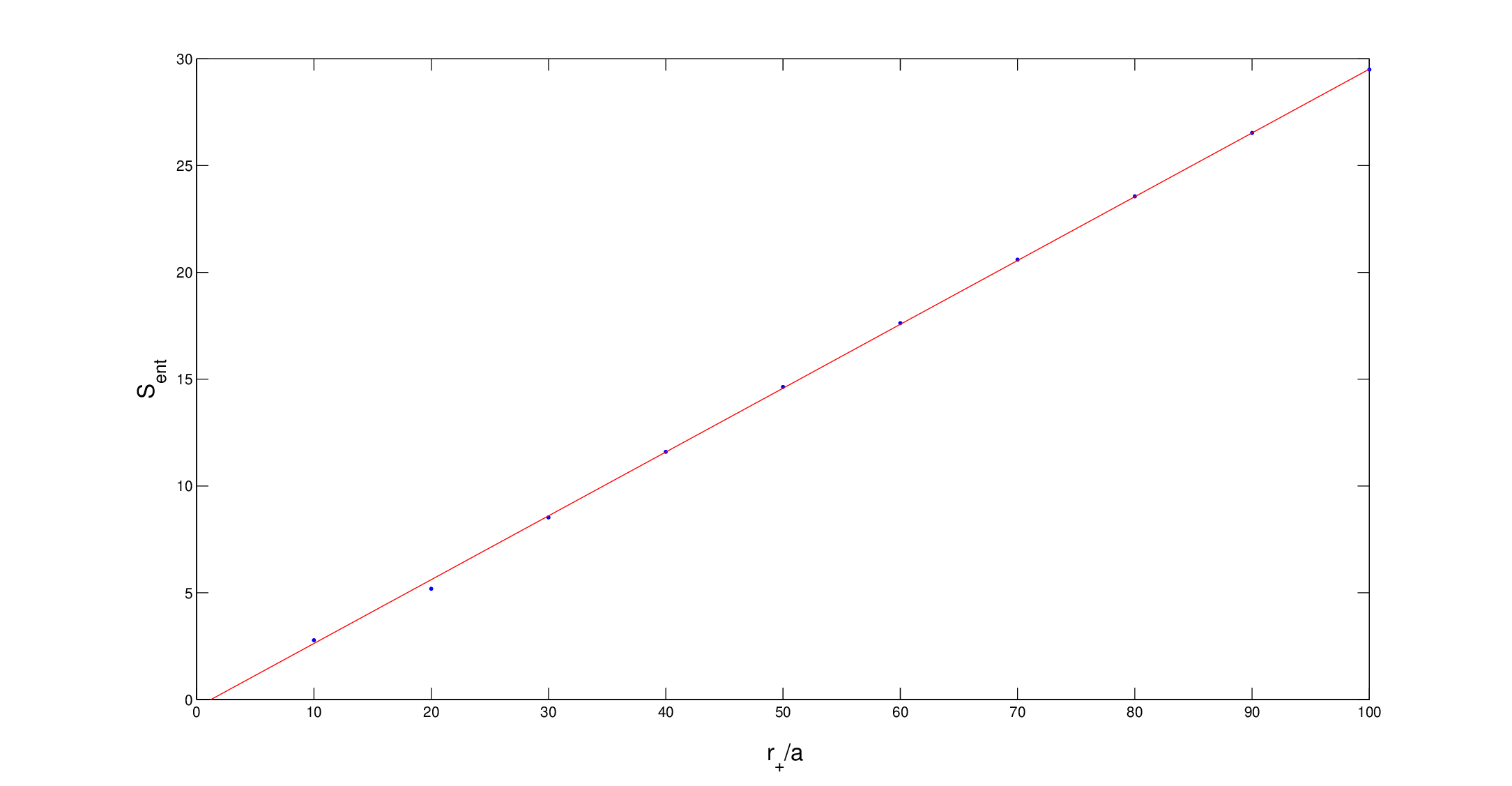}
\caption{Entanglement entropy $S_{ent}$ for $J$ =0 is shown as a functions of $n$ for N=100.}
\label{the-label-for-cross-referencing}
\end{figure}
\begin{figure}
\centering
\includegraphics[width=.75\textwidth]{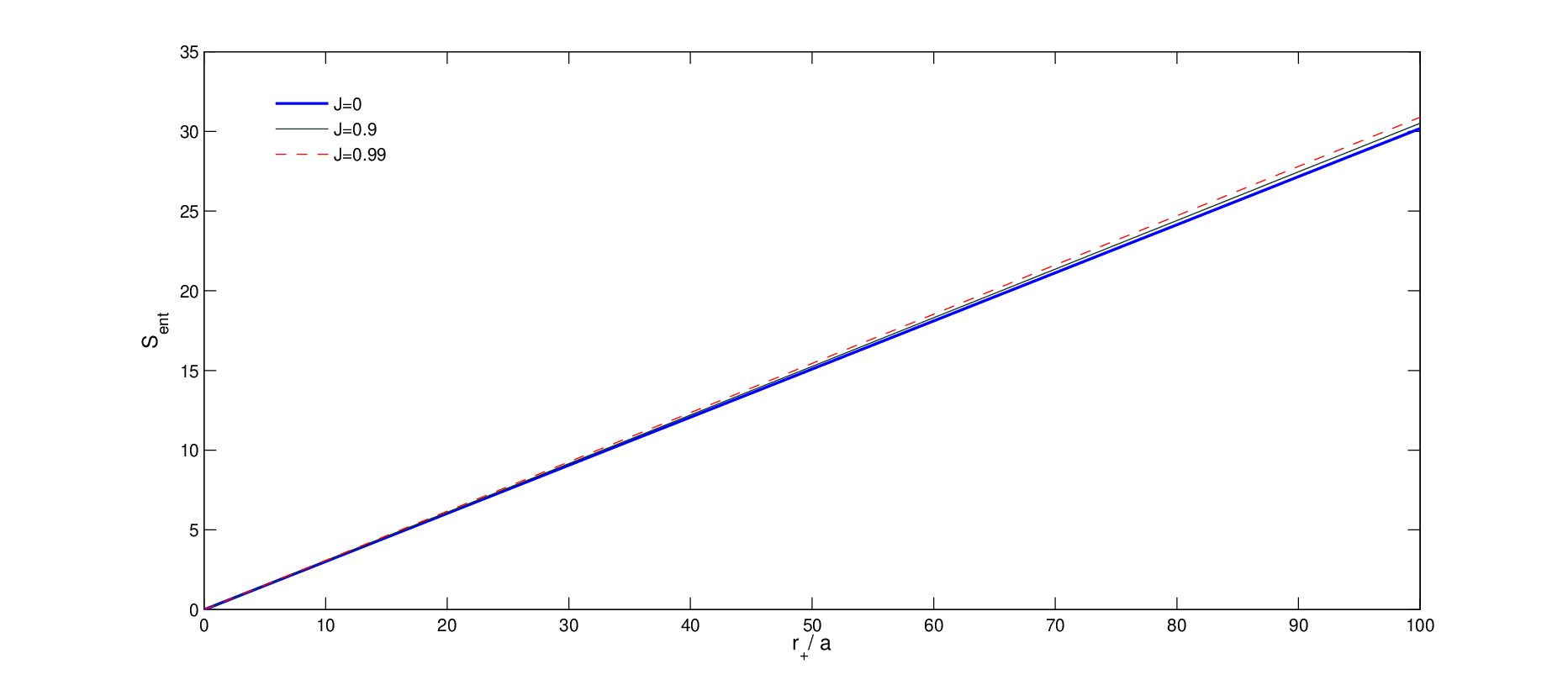}
\caption{Entanglement entropy $S_{ent}$ is shown as a functions of $n$ for N=100 for different values of $J$ =0,0.9 and 0.99.}
\label{the-label-for-cross-referencing}
\end{figure}
 
\section{Results and Discussions}
 In this paper, we have calculated the entanglement entropy of quantum scalar fields in BTZ black hole spacetime. The model is very similar to a bunch of harmonic oscillators for free fields in curved spacetime and seem to capture the area law of black hole thermodynamics. We have also calculated the leading and sub-leading terms of the entropy. The sub-leading term denotes the first quantum correction of the entropy of black holes and we has also estimated the co-efficent of this term using logarithimic fitting to our data points. The result seem to agree with that of Mann and Solodukhin \cite{MS}.
 
These results can also be compared qualitatively with entanglement entropy of Schwarzschild black hole and Kerr-Newman black hole spacetime in four dimensions \cite{Mukohyama:1998C}. These result can be generalized in the case of charged and charged rotating black holes in (2+1)-dimensions. Cadoni and Melis \cite{Cadoni:2009} have evaluated holographic entanglement entropy of BTZ black hole and seem to have an exact expression for the entropy.  It would be interesting to  compare our result with that from holographic entanglement entropy approach and find the relation, if any.

\section*{Acknowledgments}
The work of Dharm Veer Singh is supported by Rajiv Gandhi National Fellowship Scheme University Grant Commission (Under the fellowship award no. F.14-2(SC)/2008 (SA-III)) of Government of India.

\end {document}